\documentclass[preprint,nofootinbib,floats,eqsecnum,showpacs,11pt]{revtex4-1}
\usepackage{amsfonts}
\usepackage{amsmath}
\usepackage{amssymb}
\usepackage{color}
\usepackage{graphicx}
\usepackage{braket} 
\usepackage[font=small,format=plain,labelfont=bf,up,textfont=it,up]{caption}
\makeatletter
\@addtoreset{equation}{section} 
\makeatother

\def\appendix#1
{
\addtocounter{section}{1}\setcounter{equation}{0}
 \renewcommand{\thesection}{\Alph{section}}
 \section*{
 \thesection\protect\indent \parbox[t]{16.715cm}{#1}}
 \addcontentsline{toc}{section}{Appendix \thesection\ \ \ #1}
}

\begin{document}

\title{Quantum Reference Frames via Transition Amplitudes in Timeless Quantum Gravity}

\author{Jerzy Lewandowski} \email{jerzy.lewandowski@fuw.edu.pl }
\author{Chun-Yen Lin} \email{cynlin@ucdavis.edu }
\affiliation{Faculty of Physics, University of Warsaw, Pasteura 5, 02-093 Warsaw, Poland
}

\begin{abstract}
\textbf{We propose an algorithm of extracting Schr\"odinger theories under all viable physical time from the Einstein-Hilbert path integral, formulated as the timeless transition amplitudes $\hat{\mathbb{P}}:\mathbb{K} \to \mathbb{K}^*$ between the boundary states in a kinematic Hilbert space $\mathbb{K}$. Each of these Schr\"odinger theories refers to a certain set of quantum degrees of freedom in $\mathbb{K}$ as a background, with their given values specifying moments of the physical time. Restricted to these specified background values, the relevant elements of $\hat{\mathbb{P}}$ are transformed by the algorithm into the unitary propagator of a corresponding reduced phase space Schr\"odinger theory. The algorithm embodies the fundamental principle of quantum Cauchy surfaces, such that all the derived Schr\"odinger theories emerge from one timeless canonical theory defined by $\hat{\mathbb{P}}$ as a rigging map, via the relational Dirac observables referring to the corresponding backgrounds. We demonstrate its application to a FRW loop quantum cosmological model with a massless Klein-Gordon scalar field. Recovering the famous singularity-free quantum gravitational dynamics with the background of the scalar field, we also obtain in another reference frame a modified Klein-Gordon field quantum dynamics with the background of the spatial (quantum) geometry.}
\end{abstract}

\maketitle

\section{Introduction}

The Fadeev-Popov path integral of the Einstein-Hilbert action plays a conceptually central role in background-independent quantum gravity, for both the covariant \cite{refqg2}\cite{refqg3}\cite{ref3} and canonical \cite{refqg0}\cite{refqg1} formulations. 

Remarkably, recent works \cite{projector1}\cite{projector2}\cite{projector3} have reached beyond formal definitions of this path integral, prescribing it rigorously in either calculable analytic forms or perturbation expansions. In such exact prescriptions, the integral becomes an operator $\hat{\mathbb{P}}: \mathbb{K} \to \mathbb{K}^*$ from a kinematic Hilbert space $\mathbb{K}$ to its algebraic dual space. It has Hermitian matrix elements $\mathbb{P}\big(\ket{\psi_f},\ket{\psi_i}\big)\equiv\braket{\hat{\mathbb{P}}\cdot{\psi}_f|{\psi}_i}$ giving the value of the integral between the boundary states $\psi_f,\psi_i \in \mathbb{K}$. It satisfies the quantum constraint equations $\mathbb{P}\big(\hat{C}_{\mu}\ket{\psi_f},\ket{\psi_i}\big)=\mathbb{P}\big(\ket{\psi_f},\hat{C}_{\mu}\ket{\psi_i}\big)=0$ imposed by a set of quantum constraint (field) operators $\{ \hat{C}_{\mu}:\mathbb{K} \to \mathbb{K} \}$ representing the system of scalar and momentum constraints in canonical general relativity. Particularly, the constraint equation corresponding to the scalar constraints $\hat{C}_{0}$ is known as the Wheeler-DeWitt equations \cite{WDW}. In both of the covariant and canonical formulations, the unitary quantum dynamics is expected to emerge from this operator $\hat{\mathbb{P}}$.

 Based on the previous line of works \cite{CY0}\cite{CY1}\cite{CY2}, this paper is devoted to finding a fundamental relation between the unitary dynamics and the elements of $\hat{\mathbb{P}}$, the precise values of which are becoming accessible to us. Let us now introduce such an idea by reviewing the roles of $\hat{\mathbb P}$ in the two formulations. 

In the covariant formulation \cite{refqg2}\cite{refqg3}\cite{ref3}, the operator $\hat{\mathbb{P}}$ plays the role of the ``superspace transition amplitudes" \cite{WDW2}, which differ from the usual propagators in a given physical time. In the quantization of a usual Hamiltonian theory with a fixed background spacetime decoupled from the dynamical fields, the corresponding path integral would satisfy a Schr\"odinger equation and yield the unitary propagator for the dynamics. This is not the case with the path integral of the Einstein-Hilbert action \cite{barvinsky}, which is for the fully interacting gravity and matter fields. Without a fixed background spacetime, the notion of physical time here has to come from a chosen proper set of the interacting degrees of freedom in $\mathbb{K}$ serving as a relational background.  This background would be represented by a set of operators we denote as $\{\hat{T}_{\mu}\}$. Accordingly, the rest of the degrees of freedom are then treated as dynamical and are represented by a set of operators we denote as $\{\hat{X}_I\}$ (conjugate to $\{\hat{P}_I\}$). A physical time $t$ may be fully specified by referring to the background values at each moment as a set of given functions $\{{T}_{\mu}(t)\}$. This way, we expect the evolution in along $t$ to be given by the relevant elements $\mathbb{P}\big(\ket{{T}_{\mu}(t'),{X}'_I},\ket{{T}_{\mu}(t),{X}_I}\big)$, as a transformation in the $X_I$ sector of $\mathbb{K}$. However, according to the Wheeler-DeWitt equation, this transformation turns out to be nonunitary \cite{barvinsky} except for very few special cases. On the other hand, if $\{{T}_{\mu}(t)\}$ properly gauge fixes a sector of the physical states, such that they become coordinatized by $({X}_I, P_I)$ at each $t$, it is known that general relativity in this sector can be formulated as a usual Hamiltonian theory in the reduced phase space $({X}_I, P_I)$ under one fixed notion of physical time $t$. In the reduced phase space quantization approach \cite{reduced1}, one simply quantizes this Hamiltonian theory with the physical time specified at classical level. There exist many studies \cite{barvinsky}\cite{reflqc2}\cite{reflqc0} relating the Fadeev-Popov path integral $\mathbb{P}\big(\ket{{T}_{\mu}(t'),{X}'_I},\ket{{T}_{\mu}(t),{X}_I}\big)$ to the corresponding reduced phase space path integral which gives a unitary propagator. From detailed semiclassical analysis \cite{barvinsky}\cite{barvinsky2}, it is known that the former may be converted into the latter through multiplying the former's integrand by certain boundary factors, in the form of path functionals $\Lambda({X}'_I,{P}'_I; t')$ and $\Lambda({X}_I,{P}_I; t)$ depending only on the end point variables at $t'$ and $t$. A desirable goal in this line is to go beyond semiclassical level, for a general rule of such conversions, which would then generate from $\hat{\mathbb{P}}$ exact Schr\"odinger theories in the various choices of the physical time.

In the canonical formulation via the refined algebraic quantization \cite{refqg0}\cite{refqg1}\cite{ref2}, $\hat{\mathbb{P}}$ is called the rigging map and it serves as a generalized kernel projector for the quantum constraints $\{ \hat{C}_{\mu}\}$. Its image $\mathbb{H}\subset \mathbb{K}^*$ gives the physical Hilbert space, and its elements naturally define the physical inner product in $\mathbb{H}$. Explicitly, the inner product between two physical states $\{\,|\Psi_1)\equiv \hat{\mathbb{P}}\ket{\psi_1}, |\Psi_2) \equiv \hat{\mathbb{P}}\ket{\psi_2} \, \}\subset \mathbb{H}$ is defined by \cite{ref2}
\begin{eqnarray}
\label{inner product}
(\Psi_1|\Psi_2)\equiv \mathbb{P}\big(\ket{\psi_1},\ket{\psi_2}\big).
\end{eqnarray} 
Note that the dynamics has to emerge from these physical states which are constructed without any notion of time \cite{kuchartorre}. Perhaps one of the most predominant and persuasive methods for dealing with this problem is through the relational observables \cite{kuchar}\cite{torre1}\cite{relational1}\cite{relational2}\cite{relational3}. In a proper region of the classical phase space where the mentioned reduced phase space theory can be defined, one can define the gauge-invariant phase space functions $\{{X}_I(T_{\mu}(t))\}$ as the relational observables having the values of $\{{X}_I\}$ taken at the point with $T_{\mu}=T_{\mu}(t)$ in each constraint orbit. If treated correctly, the dynamics described by $\{{X}_I(T_{\mu}(t))\}$ should be the same \cite{reduced1} as the one given by the reduced phase space Hamiltonian theory. For our timeless canonical quantum theory, one may expect the quantum dynamics to be given by the quantized relational observables $\widehat{{X}_I(T_{\mu}(t))}: \mathbb{H} \to \mathbb{H}$. Here, the difficult task is in the explicit construction of these quantum observables. At the classical level, the relational observables are highly nonlinear in the phase space coordinates, and so their quantization faces complicated ambiguities. The existing literature \cite{relational3}\cite{reflqc2} for constructing these quantum observables mainly follows the two guidelines to constrain the ambiguities, namely the requirements of the commutativity with $\{ \hat{C}_{\mu}\}$ and the self-adjointness in $\mathbb{H}$. While much progress has been achieved in the existing works, we note that the ``elementary" relational observables for a Schr\"odinger theory demand stronger properties; particularly, $\{\widehat{{X}_I(T_{\mu}(t))}\}$ together with $\{\widehat{{P}_I(T_{\mu}(t))}\}$ have to satisfy a time-independent elementary algebra, in order to provide the fixed complete set for a Schr\"odinger theory. Such stronger demands call for a deep quantum construction principle for the relational observables.

We now observe that the covariant and canonical formulations compliment  each other in the described perspectives, such that the problem of unitary evolution can be resolved by reconciling the two formulations. In the canonical formulation, for the moment, assume that for each value of $t$ we have $\{\widehat{{X}_I(T_{\mu}(t))}\}$ as a complete set of relational observables for some subspace $\mathbb{D}\subset \mathbb{H}$ so that we have an orthonormal eigenbasis $\{|{X}_I(T_{\mu}(t))\}$ for $\mathbb{D}$. Recalling that $\widehat{{X}_I(T_{\mu}(t))}$ represents $X_I$ taken at the moment when the background has the values $\{{T}_{\mu}(t)\}$, we naturally expect its eigenstate to be of the form $\{|{X}_I(T_{\mu}(t))\equiv \hat{\mathbb{P}} \hat{\Lambda}\ket{X_I,{T}_{\mu}(t)}\}$, where the possible correction $\hat{\Lambda} \equiv \hat{\Lambda}(\hat{X}_i, \hat{P}_i, \hat{T}_{\mu} )$ would commute with $\hat{T}_{\mu}$ so as not to disturb the specified value of the background. Since by assumption $\{|{X}_i(T_{\mu}(t))\}$ of each $t$ gives an orthonarmal basis for $\mathbb{D}$, we immediately obtain a unitary evolution (in a generalized Heisenberg picture) describing any physical state $\Psi\in \mathbb{D}$. The elements of this unitary propagator according to \eqref{inner product} are given by $\mathbb{P}\big(\hat{\Lambda}\ket{{T}_{\mu}(t'),{X}'_I},\hat{\Lambda}\ket{{T}_{\mu}(t),{X}_I}\big)$.  From our previous discussion, this propagator should be obtainable from the reduced phase space path integrals for the chosen background. Indeed, it agrees with the mentioned semiclassical relation with the corresponding Fadeev-Popov path integral $\mathbb{P}\big(\ket{{T}_{\mu}(t'),{X}'_I},\ket{{T}_{\mu}(t),{X}_I}\big)$; only here, the boundary factors are represented by the exact operators $\hat{\Lambda}$. From this point of view, obtaining the Schr\"odinger propagator in the covariant formulation may be equivalent to finding the elementary relational Dirac observables in the canonical formulation. They would both be solving for the same operator $\hat{\Lambda}$. 

In this paper, we present a calculation algorithm for this framework and demonstrate its fundamental principle. Under a choice of $\{\hat{T}_{\mu}\}$ and $\{{T}_{\mu}(t)\}$, the input of the algorithm is just the relevant elements $\{\mathbb{P}\big(\ket{{T}_{\mu}(t'),{X}'_I},\ket{{T}_{\mu}(t),{X}_I}\big)\}$ assumed to be calculable. The output of the algorithm gives us (1) whether the specified background provides a valid notion of time and, (2) if yes, the space $\mathbb{D}\subset\mathbb{H}$ and the Schr\"odinger wave functions $\Psi[{X}_I](t)$ describing the evolution of each $\Psi\in \mathbb{D}$ in this physical time. Particularly, the output 1 contains the formulation of the exact criteria of a physical time at the deep quantum-level, and the output 2 implicitly determines the exact forms of the observables $\{\widehat{{X}_I(T_{\mu}(t))},\widehat{{P}_I(T_{\mu}(t))}\}$ and of the physical Hamiltonian. Subsequently, the outputs also yield transformations between two Schr\"odinger representations of the same physical state. From these, we obtain the notion of variable quantum reference frames.

We will illustrate this algorithm by applying it to the well-known FRW loop quantum cosmology \cite{improvedlqc}\cite{reviewlqc} with a massless Klein-Gordon scalar field. We will first derive the famous singularity-free quantum evolution of gravity in the scalar field background; then, for a new result we will show that the same model also gives quantum dynamics of the Klein-Gordon field in the background of the quantum spatial geometry. In their overlaping domain, the two dynamics are equivalent through a transformation of the quantum reference frames.

\section{ Calculation Algorithm and Underlying Principle}

Suppose we have the transition amplitudes $\hat{\mathbb{P}}: \mathbb{K} \to \mathbb{K}^*$ that solves the quantum constraints $\{\hat{C}_{\mu}\}$ and we denote a complete set of self-adjoint operators for $\mathbb{K}$ as $\{\hat{X}_i\}$ with their conjugate momenta being $\{\hat{P}_i\}$. 

Depending on the type of dynamics one wishes to describe, one first divides the degrees of freedom in $\mathbb{K}$ into two commuting sectors: a background sector and a dynamical sector. Explicitly, if one wishes to derive the quantum dynamics of the sector $\{ (\hat{X}_{I}, \hat{P}_{I})\} \subset \{ ( \hat{X}_{i}, \hat{P}_{i})\} $, one is led to choose the remaining set $\{ (\hat{X}_{\mu}, \hat{P}_{\mu})\} \equiv  \{ ( \hat{X}_{i}, \hat{P}_{i})\} - \{ (\hat{X}_{I}, \hat{P}_{I})\} $ as the background sector. A physical time $t$ defined using this background sector is specified by a chosen set of background field operators $\{\hat{T}_{\mu}= \hat{T}_{\mu}(\hat{X}_{\nu}, \hat{P}_{\nu})\}$ and a given set of functions $\{{T}_{\mu}(t)\}$ over $t$ taking values in the spectra of $\{\hat{T}_{\mu}\}$. The set of functions $\{{T}_{\mu}(t)\}$ then specifies a one-parameter family of kinematic eigenspaces $\{\mathbb{S}^{{t}}\}\equiv Span\{\ket{{T}_{\mu}(t), X_I}\}$, among which the relevant transition amplitudes are to be taken.

To extract the dynamics in this background time, we introduce a general $\mathbb{S}^{{t}}$-preserving operator $\hat{\Lambda}(\hat{X}_I,\hat{P}_I,\hat{T}_\mu): \mathbb{S}^{{t}}\to \mathbb{S}^{{t}}$ and look for the assumed reduced phase space propagator $$\mathcal{T}_{t',t}({X}^{'\Lambda}_I,{X}^{\Lambda}_I)\equiv \mathbb{P}\big(\hat{\Lambda}\ket{{T}_{\mu}(t'),{X}'_I},\hat{\Lambda}\ket{{T}_{\mu}(t),{X}_I}\big).$$ For more transparent expressions, we introduce three square matrices $\mathbb{P}_{t',t}$, $\Lambda_t$ and $K$ with their matrix elements defined by
\begin{eqnarray}
 \mathbb{P}_{t',t}({X}'_I, {X}_I)\equiv\mathbb{P}\big(\ket{{T}_{\mu}(t'),{X}'_I},\ket{{T}_{\mu}(t),{X}_I}\big)\,,\nonumber\\ \,\,\Lambda_t({X}'_I,{X}_I)\equiv\bra{{T}_{\mu}(t),{X}'_I}\hat{\Lambda}\ket{{T}_{\mu}(t),{X}_I} \,\,\text{and}\,\, K({X}'_I,{X}_I)\equiv\braket{{X}'_I|{X}_I}. \nonumber
\end{eqnarray}
Here, the $K$ represents the dynamical part of the factorized kinematic inner product $\braket{{X}'_I,{T}'_{\mu}|{X}_I,{T}_{\mu}}\equiv\braket{{X}'_I|{X}_I} \braket{{T}'_{\mu}|{T}_{\mu}}$. Using these matrix notations, we perform the following steps:

\begin{enumerate}
\item Calculate the relevant transition amplitudes $\{\mathbb{P}_{t',t}({X}'_I, {X}_I)\}$ with arbitrary $t'$ and $t$.
\item Use the $\mathbb{P}_{t,t}$ to solve for a $\hat{\Lambda}$ satisfying $\mathcal{T}_{t,t}({X}^{'\Lambda}_I,{X}^{\Lambda}_I)= \braket{{X}'_I|{X}_I}$, which in our matrix notation can be written as 
\begin{eqnarray}
\label{inner product3}
\Lambda^\dagger_t \,\,\mathbb{P}_{t,t} \,\,\Lambda_t = K.
\end{eqnarray}
A specific solution $\Lambda_t$ for every $t$ then yields one specific solution for $\hat{\Lambda}$. If there is no solution for $\Lambda_t$, the $t$ cannot be a physical time and the assumed Schr\"odinger theory does not exist. 
\item Use a found solution $\hat{\Lambda}$ and check if the assumed propagator $\mathcal{T}_{t_1,t_2}({X}^{'\Lambda}_I,{X}^{\Lambda}_I)$ gives a unitary evolution along $t$ in the dynamical sector of $\mathbb{K}$, that is, whether 
\begin{eqnarray}
\label{inner product4}
\big(\Lambda^\dagger_{t'} \,\,\mathbb{P}_{t',t} \,\,\Lambda_t\big)^\dagger \cdot\, \big(\Lambda^\dagger_{t'} \,\,\mathbb{P}_{t',t} \,\,\Lambda_t \big)= I
\end{eqnarray}
holds for the matrices. When the above unitarity condition is not satisfied, the assumed Schr\"odinger theory does not exist and the $t$ cannot be a physical time. When the unitarity is satisfied, one obtains the reduced phase space propagator of a Schr\"odinger theory with the physical time $t$, with $\{(\hat{X}_I, \hat{P}_I)\}$ as a conjugate pair of complete set observables. One may then derive from the propagator the physical Hamiltonian in terms of $\{(\hat{X}_I, \hat{P}_I)\}$.
\end{enumerate} 

We now show that all the Schr\"odinger theories obtained through the above algorithm are unified in a single (generalized) Heisenberg picture. This Heisenberg picture is provided by the Dirac theory constructed through the refined algebraic quantization procedure, with $\hat{\mathbb{P}}: \mathbb{K} \to \mathbb{H}\subset \mathbb{K}^*$ serving as the rigging map. 

The images of $\{\mathbb{S}^{{t}}\}$ under the rigging map correspond to a family of physical subspaces $\{\,\mathbb{D}_t \equiv Image[\,\hat{\mathbb{P}}|_{{\mathbb{S}}^{t}}\,] \,\}$. It can be shown easily that $\mathbb{S}^{{t_1}}$ is in one-to-one correspondence with $\mathbb{D}_{t_1}$ under the map $\hat{\mathbb{P}}|_{{\mathbb{S}}^{t_1}}$, if and only if a solution for $\Lambda_{t_1}$ exists. When this happens, there is a right-inverse map $\hat{\Pi}_{t_1}:\mathbb{D}_{t_1} \to \mathbb{S}^{{t_1}}$ satisfying $\hat{\mathbb{P}}\,\hat{\Pi}_{t_1}= \hat{I}$, called a quantum Cauchy surface. Clearly, the quantum Cauchy surface $\hat{\Pi}_{t_1}$ represents each physical state of the quantum spacetime in $\mathbb{D}_{t_1}$ with a unique element in $\mathbb{S}^{t_1}$, which describes the spatial slice of the quantum spacetime where the background fields take the values $\{{T}_{\mu}(t_1)\}$. When such a solution does not exist, the map $\hat{\mathbb{P}}|_{{\mathbb{S}}^{t_1}}$ is degenerate and the background sector needs to be extended to provide sufficient gauge fixing conditions. 

For each quantum Cauchy surface, our algorithm also identifies a special isometry map $\hat{\mathbb{P}}\,\hat{\Lambda}:\mathbb{S}^{t_1}\to \mathbb{D}_{t_1}$ which preserves the inner products in the two Hilbert spaces. Through this isometry, the self-adjoint complete sets $\{\hat{X}_{I},\hat{P}_{I}\}$ in $\mathbb{S}^{{t_1}}$ naturally induce the corresponding self-adjoint complete sets of Dirac observables $\{\hat{X}^{\Lambda}_{I}(t_1),\hat{P}^{\Lambda}_{I}(t_1)\}$ in $\mathbb{D}_{t_1}$. These complete sets of Heisenberg observables in $\mathbb{D}_{t}$ induced by the quantum Cauchy surfaces take the explicit form of 
\begin{eqnarray}
\label{Dirac observ 0}
(\,\hat{{X}}_I^{{\Lambda}}(t)\,,\,\hat{{P}}_I^{{\Lambda}}(t)\,) \equiv \hat{\mathbb{P}}\,{\Lambda}\,(\,\hat{{X}}_I\,, \,\hat{{P}}_I\,)\,{\Lambda}^{\!-\!1}\,\hat{\Pi}_{t} \,:\, \mathbb{D}_{t} \to \mathbb{D}_{t}
\end{eqnarray}
This desirable form is such that 
\begin{eqnarray}
\label{Dirac observ 1}
 F(\hat{{X}}_I^{{\Lambda}}(t),\hat{{P}}_I^{{\Lambda}}(t))=F(\hat{{X}}_I,\hat{{P}}_I)^{{\Lambda}}(t)\,\, \,\text{and} \,\,\,  [\,\hat{{X}}_I^{{\Lambda}}(t)\,,\,\hat{{P}}_I^{{\Lambda}}(t)\,]=[\,\hat{{X}}_I\,, \,\hat{{P}}_I\,]^{{\Lambda}}(t)\,,
\end{eqnarray}
thus the observables have the algebra and spectrum  identical to their kinematic counterparts'. Further, an orthonormal eigenbasis for $\mathbb{D}_{t}$ of these Dirac observables, satisfying $\hat{{X}}^{{\Lambda}}_I(t)|{X}^{{\Lambda}}_I(t)\,)={X}_I\,|{X}^{{\Lambda}}_I(t)\,)$ and $(\,{X}^{'{\Lambda}}_I(t)|{X}^{{\Lambda}}_I(t)\,)= \braket{{X}'_I|{X}_I}$, is given by
\begin{eqnarray}
\label{eigenbasis}
 |{X}^{{\Lambda}}_I(t)\,) \equiv \hat{\mathbb{P}}\,{\Lambda} \ket{{T}_{\mu}(t), {X}_I}. 
\end{eqnarray}
Therefore, we have
\begin{eqnarray}
\label{inner product5}
\mathcal{T}_{t_2,t_1}({X}^{'\Lambda}_I,{X}^{\Lambda}_I)= (\,{X}^{'{\Lambda}}_I(t_2)|{X}^{{\Lambda}}_I(t_1)\,).
\end{eqnarray}

In the last step of the algorithm, we examine the unitarity of the matrices $(\,{X}^{'{\Lambda}}_I(t_2)|{X}^{{\Lambda}}_I(t_1)\,)$. Now it is clear that this unitarity implies $\mathbb{D}_{t_1}=\mathbb{D}_{t_2}\equiv \mathbb{D}$. When this happens the space $\mathbb{D}$ provides the generalized Heisenberg state space for the physical time $t$. This unitarity can thus be viewed as a global hyperbolicity condition for the quantum spacetimes in $\mathbb{D}$, realized by the foliation of the quantum Cauchy surfaces associated with $t$. Each physical state $\Psi_{\mathbb{D}}\in \mathbb{D}$ is described by a Schr\"odinger wave function $\Psi_{\mathbb{D}}[{X}^{{\Lambda}}_I](t)\equiv (\,{X}^{{\Lambda}}_I(t)|\Psi_{\mathbb{D}}\,)$ evolving in the physical time $t$ with the propagator given by \eqref{inner product4}. Lastly, a breakdown of the unitarity in $(\,{X}^{'{\Lambda}}_I(t_2)|{X}^{{\Lambda}}_I(t_1)\,)$ implies $\mathbb{D}_{t_1}\neq\mathbb{D}_{t_2}$ and that the quantum Cauchy surfaces along $t$ do not provide a global hyperbolic foliation to a fixed set of quantum spacetimes.

The full set of solutions for \eqref{inner product3} and \eqref{inner product4} is given by the ``left-unitary" class of any given solution $\hat{\Lambda}$; the set is generated by $\hat{\Lambda}'=\hat{U} \hat{\Lambda}$ with an arbitrary left-unitary transformation $\hat{U}(\hat{X}_I, \hat{P}_I, \hat{T}_\mu): {\mathbb{S}}^{t} \to {\mathbb{S}}^{t} $. These transformations really are the canonical transformations in a usual quantum theory. To see this, suppose we switch from using the $\hat{\Lambda}$ to using the $\hat{\Lambda}'$ for our observables \eqref{Dirac observ 0}. It follows from \eqref{Dirac observ 1} that the replacement is equivalent to a unitary redefinition of observables, because we have
$$\big(\,\hat{{X}}_I^{{\Lambda'}}(t)\,, \,\hat{{P}}_I^{{\Lambda'}}(t)\,\big)= \hat{U}^{\dagger}(\hat{X}_I^{\Lambda}(t), \hat{P}_I^{\Lambda}(t), T_\mu(t))\,\,\big(\,\hat{{X}}_I^{\Lambda}(t) \,,\,\hat{{P}}_I^{\Lambda}(t)\,\big)\,\,\hat{U}(\hat{X}_I^{\Lambda}(t), \hat{P}_I^{\Lambda}(t), T_\mu(t)).$$ The real task here is to understand the physical meaning of $(\,\hat{{X}}_I^{{\Lambda}}(t)\,,\,\hat{{P}}_I^{{\Lambda}}(t)\,)$ given by a particular $\hat{\Lambda}$, such as the classical limits of these observables.

Indeed, our algorithm is based on quantum-level considerations; in order to find a Schr\"odinger theory with a certain desired classical limit, we need to choose the background fields $\{\hat{T}_\mu\}$ such that the corresponding observables \eqref{Dirac observ 0} can truly represent the needed classical relational observables. Such choice of the background fields requires instructional guidance based on classical intuitions. Here, let us put forth one such guidance coming from the following simple consideration. Suppose the Schr\"odinger theory with the physical time $t$ discussed above has a classical limit of a Hamiltonian dynamics in the reduced phase space $\{({X}_I, {P}_I)\}$, under the assigned background value $T_{\mu}(t)$. This means at any moment $t_1$ the observable's values $({X}_I, {P}_I)(t_1)$ together with the background value $T_{\mu}(t_1)$ must correspond to a unique point $({X}_I, {P}_I,{X}_\mu, {P}_\mu)(t_1)$ on the constraint surface in the original phase space, thereby resolving the well-known Gribov ambiguity. Suppose we have specifically chosen the frame in which $\{{X}_\mu\}$ has certain specified values $\{{X}_\mu(t)\}$; then, our consideration implies that the background field should be chosen as $T_{\mu}({X}_\nu, {P}_\nu)= \Theta(g({X}_\nu, {P}_\nu))\,{X}_\mu $. Here, the $\Theta(x)$ is the Heaviside step function, and $g({X}_\nu, {P}_\nu)$ is a phase space function satisfying $g>0$ on the constraint surface only in the region where the mentioned one-to-one correspondence holds. Clearly, the proper function $g({X}_\nu, {P}_\nu)$ can be determined by studying the constraints $\{C_\mu\}$, and this way, the condition $T_{\mu}(t)=t\neq 0$ excludes automatically the regions of the constraint surface containing the additional Gribov copies. We are thus instructed to choose our quantum background fields to be $\hat{T}_{\mu}=\Theta(\hat{g}(\hat{X}_\nu, \hat{P}_\nu))\,\hat{X}_\mu +O(\hbar)$. Note that the above consideration is based on the uniqueness of the dynamical trajectory, a form of the ``classical unitarity" condition. Thus, we expect the quantum unitarity \eqref{inner product4} imposed by our algorithm to agree with this choice of $\hat{T}_{\mu}$ up to an error of $O(\hbar)$, and then the exact quantum unitarity should serve to further constrain the form of $\hat{T}_{\mu}$ in the quantum-level.

For the Schr\"odinger dynamics of a different set of fields $\{({\mathcal{X}}_{I},{\mathcal{P}}_{I})\}$, we would use the physical time based on another complete set decomposition $\{({\mathcal{X}}_{i},{\mathcal{P}}_{i})\}\equiv \{({\mathcal{X}}_{I},\mathcal{{P}}_{I})\} \cup \{({\mathcal{X}}_{\mu},{\mathcal{P}}_{\mu})\}$ for $\mathbb{K}$. Again, we choose the reference frame by choosing the background fields $\{\hat{\mathcal{T}}_{\mu} (\hat{\mathcal{X}}_{\mu},\hat{\mathcal{P}}_{\mu})\}$ with the values $\{{\mathcal{T}}_{\mu}(\tau)\}$ over the physical time $\tau$. This may lead to another Schr\"odinger theory describing the physical states in ${\mathbb{D}'}\subset \mathbb{H}$. Then, any physical state $\Psi \in {\mathbb{D}'} \cap \mathbb{D} $ can be described in both reference frames and represented as either $\Psi[{X}_I]^{{\Lambda}}(t)$ or $\Psi[\mathcal{X}_I]^{{\bar\Lambda}}(\tau)$. The transformation between the two wave functions is the transformation between the two quantum reference frames, associated with the two families of quantum Cauchy surfaces.

Lastly, we note that the background field operators $\hat{T}_{\mu}$ coordinatize a physical time with the specified values $T_\mu(t)$. They thus play a role fundamentally different from that of the operators $\hat{{X}}_I$, which represent the dynamical observables at each moment of time. Consequently, these operators are not required to be self-adjoint, but instead they have to satisfy the condition of capturing a true physical time. Concretely, the condition is the solvability of Eqs. \eqref{inner product3} and \eqref{inner product4} for the $\hat{\Lambda}$. Indeed, we have seen its physical meaning: the background field values must specify quantum Cauchy surfaces that provide globally hyperbolic foliations to the set of quantum spacetimes fluctuating in the dynamical degrees of freedom.

\section{ Reference Frames in Spatially Flat FRW Loop Quantum Cosmology }

We now demonstrate the algorithm by applying it to the FRW loop quantum cosmology \cite{reviewlqc}\cite{improvedlqc}-- a quantum cosmological model on the homogeneous and isotropic sector of general relativity incorporating the essential features of loop quantum gravity. We will work in the spatially flat case with zero cosmological constant, and with a minimally coupled massless Klein-Gordon scalar field. 

Restricting to the homogeneous and isotropic sector of general relativity described in the comoving frames that manifest the symmetry, we obtain our kinematic phase space canonically coordinatized by $(c,p,\phi, P_{\phi})$, with all the components taking values from $\mathbb{R}$. The gravitational sector is described by $(c,p)$, the symmetrically reduced Ashtekar variables \cite{Dirac1}, which provide the extrinsic curvature $K$ and the scale factor $a$ of space through $\gamma K= \mathcal{V}^{-1/3}c$ and $a= \mathcal{V}^{-1/3}\sqrt{|p|}$; here $\mathcal{V}$ is the coordinate volume of a chosen spatial comoving cell, and the real number $\gamma$ is known as the Barbero-Immirzi parameter \cite{Dirac1}. Note that a change in the value of $\gamma$ corresponds to a rescaling of the Ashtekar variables. For the scalar field sector, $\phi$ and $P_{\phi}$ represent the field value in the comoving space and the field's total conjugate momentum in $\mathcal{V}$. The nontrivial Poisson brackets among these variables are given by $\{c,p\}=\frac{8\pi G \gamma}{3}$ and $\{\phi, P_{\phi}\}=1$.  Under the partial gauge fixing with the comoving frames, the full diffeomorphism symmetry of general relativity is reduced to the one-dimensional diffeomorphism invariance in the comoving temporal coordinate. Consistently, the full constraint system is also simplified to just one reduced scalar constraint 
\begin{equation}
\begin{split}
\label{scalar constr}
{C}_0=\mathcal{C}_g(c,p)+\mathcal{C}_{\phi}(\phi, P_{\phi})= -\frac{6}{\gamma^2}\,c^2\,p^2 + 8 \pi G \, P^2_{\phi}.
\end{split}
\end{equation}
Here and in the following we set the speed of light to be unity. The constraint governs both the initial data and the dynamics, and predicts the initial bigbang singularity whenever the conserved $P_{\phi}$ is nonzero.

\subsection{Kinematic Setting}

Inspired by loop quantum gravity, the FRW loop quantum cosmology starts by reformulating the standard classical theory above, describing the extrinsic curvature using a holonomy variable \cite{reviewlqc}\cite{improvedlqc}
\begin{equation}
\begin{split}
\label{holonomy}
{\mathcal{N}}_{ 2\bar{\mu}}(c,p) \equiv e^{ i\bar{\mu}\,c} \,;\,\, \bar{\mu}(p)\equiv \sqrt{\frac{\Delta}{|p|}}.
\end{split}
\end{equation}
This variable carries the meaning of the integral of the extrinsic curvature along an arbitrary geodesic in the space, which has the given physical length $\sqrt{\Delta}$. The $\Delta$ represents the minimum nonzero value of the spatial area spectrum predicted by loop quantum gravity, and so this variable has the natural meaning of a holonomy over a minimal geodesic in the quantum geometry of space. The new conjugate pair for the gravitational sector is then chosen to be $({\mathcal{N}}_{2\bar{\mu}}, v)$, where $v(p)\equiv (2\pi \gamma l^2_p \sqrt{\Delta})^{-1}\text{sgn} (p)|p|^{3/2}$. This formulation provides a direct analogy to the full theory of loop quantum gravity, which then suggests treating the original constraint $\mathcal{C}_g(c,p)$ as given by $C_g(c,p)$, with $C_g(c,p)\equiv \mathcal{C}_g(\sin(\bar{\mu}c)/\bar{\mu}\,,\,p)$ mimicking the regularized scalar constraint in the full theory \cite {Dirac1} and the value of $\Delta$ set to be $\Delta\sim l^2_p$. Given that $\lim_{l_p\to 0}C_g(c,p)=\mathcal{C}_g(c,p)$, $C_g$ is chosen to be the classical scalar constraint we quantize in the FRW loop quantum cosmology.

Standard canonical quantization of the model formulated with the holonomy variable leads to our kinematic Hilbert space; in this paper we follow the specific prescription given in Ref. \cite{reviewlqc}. The holonomy variable is quantized into an excitation operator $\hat{\mathcal{N}}_{2\bar{\mu}}$ acting on the eigenstates of $\hat{v}$ as
\begin{equation}
\begin{split}
\label{holonomy action}
\hat{\mathcal{N}}_{2\bar{\mu}} \ket{v, {P}_{\phi}} = \ket{v+2, {P}_{\phi}}\,;\,\,\hat{\mathcal{N}}^\dagger_{2\bar{\mu}} \ket{v, {P}_{\phi}} = \ket{v-2, {P}_{\phi}}.
\end{split}
\end{equation}
This crucial algebra allows a superselected kinematic sector with a discretized volume spectrum characterizing the quantum spatial geometry similar to the one in the full theory. Another gravitational operator $\hat{\Omega}$ is introduced for giving $\hat{C}_g\equiv \frac{-6}{\gamma^2}\hat{\Omega}^2$, and it is defined as 
\begin{eqnarray}
\label{Omega}
\hat{\Omega}\equiv \frac{-i}{2\sqrt{\Delta}} \hat{|p|}^{3/4} \big[(\hat{\mathcal{N}}_{2\bar{\mu}}-\hat{\mathcal{N}}^\dagger_{2\bar{\mu}})\widehat{\text{sign}(p)}+\widehat{\text{sign}(p)}(\hat{\mathcal{N}}_{2\bar{\mu}}-\hat{\mathcal{N}}^\dagger_{2\bar{\mu}})\,\big]\,\, \hat{|p|}^{3/4}.
\end{eqnarray}
We also introduce the operator operator $\hat{b}$ canonically conjugate to $\hat{v}$, 
which satisfies $$\hat{b}\equiv\hbar\,\bar{\mu}(\hat{p})\hat{c}\,\,\, \text{and}\,\,\,[\hat{b}\,,\,\hat{v}]=2\hbar.$$ It has normalized eigenstates of the discrete Fourier modes over $v$ given by
 \begin{equation}
\label{b}
\begin{split}
\ket{b}=\frac{1}{\sqrt{2\pi\hbar}}\sum_v e^{-i\,b\,v/2\hbar}\ket{v}.
\end{split}
\end{equation}
Lastly, the operators in the Klein-Gordon sector are constructed in the conventional Fock representation, with $\hat{\phi}$ and $\hat{P}_{\phi}$ being, respectively, a multiplicative and a differential operator on a wave function over $\mathbb{R}$ that is the spectrum of $\hat{\phi}$.

Our kinematic Hilbert space $\mathbb{K}$ is then chosen to be a self-adjointness domain of the operators $\{\hat{p},\hat{\Omega},\hat{\phi},\hat{P}_{\phi}\}$, and it is given by\footnote{For the gravitation sector, we have chosen a superselected sector which is a positive lattice in $v$ with gaps of two units and starting from $v=1$.}   
\begin{eqnarray}
\label{space}
\mathbb{K}\equiv Span\{\, \ket{\,v=1+2n\,, \phi}\,;\,\, n \in \mathbb{Z}^+ \,,\,\phi\in \mathbb{R}\,\}
 =Span\{\ket{b, \phi}; b\in [0,2\pi\hbar]\,,\,\phi\in \mathbb{R} \}\nonumber
\\
=
 Span\{ \,\ket{\Omega, {P}_{\phi}}\,;\,\, \Omega\in \mathbb{R}\,,\,{P}_{\phi}\in \mathbb{R} \,\}, 
\end{eqnarray}
with the inner product given by 
\begin{equation}
\begin{split}
\label{kinematic inner prod}
\braket{v, \phi| v', \phi'}\equiv \delta_{v,v'} \,\delta(\phi-\phi')\,,\,\, \,\, \braket{b,\phi|b',\phi'}=\delta(b-b') \,\delta(\phi-\phi')\, \text{and}\,\,\braket{\Omega, {P}_{\phi}| \Omega', {P}'_{\phi}}\equiv \delta(\Omega-\Omega') \,\delta({P}_{\phi}-{P}'_{\phi}).
\end{split}
\end{equation}
In our algorithm's notation, we may write, for example, $\{\hat{X}_i\}= \{\hat{p},\hat{\phi}\}$ and $\{\hat{P}_i\}= \{\hat{\Omega},\hat{P}_{\phi}\}$. 

The self-adjoint quantum scalar constraint operator is constructed to be
\begin{equation}
\begin{split}
\label{grav quant scalar constr}
\hat{C}_0\equiv -\frac{6}{\gamma^2}\,\hat{\Omega}^2 + 8 \pi G \, \hat{P}^2_{\phi},\,
\end{split}
\end{equation}
and the rigging map $\mathbb{P}:\mathbb{K}\to \mathbb{K}^*$, as a precise formulation of the Fadeev-Popov path integral for this model, is given by
\begin{equation}
\begin{split}
\label{FRW riggin map}
\hat{\mathbb{P}}\equiv \int_{-\infty}^{\infty} d\lambda e^{i\lambda\hat{\mathcal{C}}}
= \delta\left(-\frac{6}{\gamma^2}\,\hat{\Omega}^2 + 8 \pi G \, \hat{P}^2_{\phi}\right).
\end{split}
\end{equation}
The ``minisuperspace" transition amplitudes given by this $\hat{\mathbb{P}}$ satisfy the symmetry-reduced Wheeler-DeWitt equations, so the path integral faces the same problem of unitary physical evolution just like in the full theory.

A prevailing approach for obtaining Schr\"odinger dynamics from Wheeler-DeWitt cosmological models is introducing the physical inner product via a conserved current operator \cite{shroedinger2}, which has zero divergence in the minisuperspace according to the reduced Wheeler-DeWitt equation. With each choice of a certain quantum variable as the background labeling the physical time, the corresponding ``time component" of the current operator can be used to construct a time-independent inner product between the solutions having proper boundary conditions. Just as in a usual quantum theory, the ambiguities of fixing the inner product in this approach are constrained by the requirements of the physical inner product: it should be Hermitian and giving non-negative norms, and it should promote the dynamical complete set in $\mathbb{K}$ into physical observables at an instant of time. The paradigmatic treatment of loop quantum cosmological models essentially adopts this approach \cite{lqc}\cite{improvedlqc}\cite{reviewlqc} while focusing on the gravitational dynamics. Particularly in our FRW setting, the scalar field has been used as the background, and the time component of the conserved current definining the inner product is just the canonical momentum of the scalar field \cite{lqc}. The obtained results, with anisotropic and inhomogeneous generalizations, show important and robust quantum gravitational corrections \cite{reviewlqc}\cite{improvedlqc} to the dynamics which dominate the early Universe and replace the initial singularity with a regular bouncing at a characteristic minimal spatial volume.

The issue of reference frames becomes relevant when we need various frames for different types of interesting quantum dynamics of the same system. In our example case, one may wish to explore the quantum dynamics of the Klein-Gordon field in the background of the quantum spacetime. For this dynamics, however, one must choose the gravitational sector to provide the background labeling the time. In the paradigmatic treatment, the new choice of physical time changes the inner product between the physical wave functions, into the one given by the new time component of the conserved current. The set of physical wave functions itself, selected from the general solutions by the new boundary conditions, may also be different. In this way, there is generally is no reference frame-independent physical Hilbert space \cite{shroedinger2} unifying the Schr\"odinger dynamics. In pursuit of the general covariance at the quantum-level, there are alternative approaches \cite{alter3}\cite{alter2}\cite{alter1} aiming to extract the dynamics under various notions of time, either from the  single timeless physical Hilbert space $\mathbb{H}$ or from the elements of $\hat{\mathbb{P}}$ as the timeless transition amplitudes.  Particularly, in the context of the Bianchi-I loop quantum cosmology model, a method \cite{alter3} that is very similar to ours has been proposed. For this particular line of ideas in quantum cosmology, our work may provide a complete generalization with a fundamental principle at the level of the full theory. 

In the following, we will use our approach to recover the existing results and also obtain the unexplored dynamics with a reference frame-independent physical Hilbert space. For this goal, we now apply our algorithm using the transition amplitudes given by \eqref{FRW riggin map} as the input.

\subsubsection{Rigging Map Matrix Elements}

With $\alpha \equiv\sqrt{48\pi G/\gamma^2}$ and $\beta\equiv \sqrt{4\pi G \gamma^2 /3}$, the rigging map \eqref{FRW riggin map} can be written as
\begin{equation}
\begin{split}
\label{FRW riggin map 2}
\hat{\mathbb{P}} 
=\int  \frac{dP_\phi}{2 \alpha {|P_\phi|}} \bigg[\,\, \ket{\Omega_{(P_{\phi})}, {P}_{\phi}}\bra{\Omega_{(P_{\phi})}, {P}_{\phi}}\,\,- \,\, \ket{-\Omega_{(P_{\phi})}, {P}_{\phi}}\bra{-\Omega_{(P_{\phi})}, {P}_{\phi}}\,\,\bigg]\,\,;\,\, \Omega_{(P_{\phi})}\equiv |\beta P_\phi|.\\
\end{split}
\end{equation}
Thus, the values of this rigging map's matrix elements in an arbitrary basis such as $\{\ket{v,\phi}\}$ or $\{\ket{b,\phi}\}$ are given by $\braket{v,\phi|\Omega,P_{\phi}}\equiv\braket{v|\Omega}\braket{\phi| {P}_{\phi}}$ or $\braket{b,\phi|\Omega,P_{\phi}}\equiv\braket{b|\Omega}\braket{\phi| {P}_{\phi}}$. They are already provided by the existing analytic calculations \cite{reviewlqc}\cite{solution} in the form  
\begin{equation}
\label{solution form0}
\begin{split}
\braket{\phi| {P}_{\phi}}= e^{i{P}_{\phi} \phi/\hbar}\,\,,\,\,\nonumber
\\
\braket{b|\Omega}=\frac{1}{\sqrt{2\pi\hbar}}\sum_v e^{i\,b\,v/2\hbar}\braket{v|\Omega}\,\,\text{and}\,\,\,
\braket{v|\pm|\Omega|}\equiv\sqrt{|\Omega|}\,\left[\,F_{|\Omega|}(v) \,\mp\, i\, G_{|\Omega|}(v) \,\right].
\end{split}
\end{equation}
where the $F_{|\Omega|}(v)$ and $G_{|\Omega|}(v)$ are real and exactly solved in Ref. \cite{reviewlqc}. 

Still, the perturbative expansions of $\braket{b|\Omega}$ and $\braket{v|\Omega}$ in terms of $\hbar$ and $l_p$ are important for understanding the results of our computations, and here we will simply take a look at the few lowest orders. First, according to  \eqref{Omega}, we have $$\hat{\Omega}=2\pi\gamma G \hbar\,\sqrt{\hat{v}}\,\sin(\frac{\hat{b}}{\hbar})\,\sqrt{\hat{v}}\,,$$ and it is straightforward to show that the normalized eigenstates of $\hat{\Omega}$ are given by 
\begin{eqnarray} 
\label{semiclass2}
\ket{\Omega}=\sqrt{\hat{v}}\,\cdot\, \frac{1}{\sqrt{\Omega}}\,\int db\,\exp\frac{-i}{\hbar} \left[\,\frac{\Omega}{4\pi\gamma G}\,\,\ln\big|\tan(\frac{b}{2\hbar})\big|\right] \ket{b}\equiv\sqrt{\hat{v}}\cdot \ket{*\Omega} .
\end{eqnarray}
Then we may find the value of $\braket{v|\Omega}=\sqrt{{v}}\braket{v|*\Omega}$ as given by
\begin{eqnarray} 
\label{semiclass3}
\braket{\Omega|v}=\sqrt{{v}}\int_{I+} d{b} \braket{*\Omega|b}\braket{b|v}+\sqrt{{v}}\int_{I_-} d{b} \braket{*\Omega|b}\braket{b|v}\,;\,\,I_\pm\equiv\{b\,;\,\, \pm\cos\frac{b}{\hbar}>0\},
\end{eqnarray} 
which can be evaluated using \eqref{b} and \eqref{semiclass2}. In doing so,
one can see that for any given $\Omega$ and $v$ there is always a pair of values $\{b_{+(\Omega,V)}\in I_+\,,\,\, b_{-(\Omega,V)} \in I_-\}$ for $b$ giving the points of the stationary phase via satisfying
\begin{eqnarray} 
\label{stationary}
\frac{\partial}{\partial b}\, \theta(b,\Omega,V)\big|_{b=b_{\pm(\Omega,V)}}&=&0\,\,, \nonumber 
\\
\text{with}\,\,\,\theta(b,\Omega,V)\equiv i\,\ln \big(\frac{\braket{\Omega|b}\braket{b|v}}{\big|\braket{\Omega|b}\braket{b|v}\big|}\big)&=&\frac{-\gamma}{3\beta^2\hbar}\left(\Omega\ln \big|\tan(\frac{b}{2\hbar})\big|-\frac{b\,V}{\sqrt{\Delta}}\right).
\end{eqnarray}
The solutions are given by the expected semiclassical relation
\begin{eqnarray}
\label{relation}
\frac{V}{\sqrt{\Delta}}\,\big|\sin\frac{b_{\pm(\Omega,V)}}{\hbar} \big|=\big|\Omega\big|\,\,.
\end{eqnarray}
Since $\partial_b\theta$ is of the order of $O(\beta^{-2}\hbar^{-1})=O(l_p^{-2})$, we may apply the stationary phase expansion to \eqref{semiclass3} and obtain  
\begin{eqnarray} 
\label{semiclass5}
\int_{I_\pm} d{b} f(b) \braket{\Omega|b}\braket{b|V}=f(b_{\pm(\Omega,V)})\,B_\pm(\Omega,V)\, e^{ -i \theta_\pm(\Omega,V) }\,\,+O( l_p^{2})\,,\ \nonumber
\\
\text{with}\,\,\,\theta_\pm(\Omega,V)\equiv\theta(b_{\pm(\Omega,V)},\Omega,V)\pm\frac{\pi}{4}\,\, \text{and}\,\,\,B_\pm(\Omega,V)\equiv\sqrt{ \frac{2\pi\,v}{\Omega\,\,|\partial_b^2\,\theta|(b_{\pm(\Omega,V)},\Omega,V) } }\,\,. \, 
\end{eqnarray} 
Here $f(b)$ can be any smooth function with $\partial_b \,f$ being of the order of $O(l_p^{0})$. This finally leads to
\begin{eqnarray} 
\label{semiclass4}
\int_{I_\pm} d{b} \braket{\Omega|b}\braket{b|V}
=\sqrt{\frac{6\pi\beta^2\hbar\,v}{\gamma\,\Omega^2\,\big|\partial^2\ln|\tan(b_\pm /2\hbar)|\big|}}\,\,\exp  i\left[\frac{\gamma}{3\beta^2\hbar}\left(\Omega\ln \big|\tan(\frac{b_\pm}{2\hbar})\big|-\frac{b_\pm\,V}{\sqrt{\Delta}}\right) \pm\frac{\pi}{4}\right] +O( l_p^{2})\,\,.
\nonumber
\\
\end{eqnarray}

Having enough control over the matrix elements of $\hat{\mathbb{P}}$, we are ready to apply our algorithm. To derive a Schr\"odinger theory for the gravitation and scalar field, we will be using, respectively, the scalar field sector and the gravitational sector as the background sector. Since there is only one scalar constraint, we expect to choose only one background field operator to define a proposed physical time.

\subsubsection{Quantum Gravity with $\hat{T}\equiv\hat{\phi}_+$}

We first study the gravitational quantum dynamics in the model using the scalar field sector as the background sector, and so we set $\{(\hat{X}_I,\hat{P}_I) \}\equiv\{(\hat{V},\hat{\Omega})\}$ and $\{(\hat{X}_{\mu},\hat{P}_{\mu}) \}\equiv \{(\hat{\phi},\hat{P}_{\phi})\}$. Following the paradigmatic setting for comparisons, we look for a Schr\"odinger theory that has the semiclassical limits of a Hamiltonian theory in the reduced phase space $({V},{\Omega})$, with the background sector satisfying $\phi(t)=t$ under the physical time $t$. 

We now use the guidance proposed earlier for choosing the background operator $\hat{T}$. Clearly, according to the form of the classical scalar constraint \eqref{scalar constr}, each given set of values to $({V},{\Omega})(t)$ and $\phi(t)=t$ corresponds to two points on the constraint surface given by the two constraint solutions $\pm P_{\phi}(t)>0$. Therefore we are instructed to use $T\equiv \Theta(P_{\phi})\,\phi\equiv {\phi}_+$. For a natural quantization of this $T$, we construct $\hat{T}(\hat{\phi},\hat{P}_{\phi})\equiv\hat{\phi}_{+}$ such that
\begin{equation}
\begin{split}
\label{time1}
\ket{{\phi}_+}\equiv \Theta (\hat{P}_\phi) \,\ket{{\phi}}\big|_{\phi=\phi_+} \,\,\,;\,\, \,\hat{\phi}_+\equiv\Theta (\hat{P}_\phi)\, \,\hat{\phi}\,\,\Theta^{\!-\!1}_L(\hat{P}_\phi)=\,\hat{\phi}\,\,\Theta (\hat{P}_\phi)\,\,+\hbar\,\delta(\hat{P}_\phi)\Theta^{\!-\!1}_L(\hat{P}_\phi)\,\,.
\end{split}
\end{equation}
Here the linear operator $\Theta^{\!-\!1}_L(\hat{x})$ is the ``left inverse" of $\Theta(\hat{x})$, defined by its operation on the basis of ``stepped Fourier modes" as 
\begin{equation}
\begin{split}
\label{inverse}
\Theta^{\!-\!1}_L(\hat{x}) \int_0^{\pm\infty} dx \,\,e^{-i px}\ket{x} \equiv \delta_{+,\pm}\,\, \int_{-\infty}^{+\infty} dx\,\, e^{-i px}\ket{x}.
\end{split}
\end{equation}

We now set $T(t)=t$ with $t$ being the proposed physical time. Introducing the abbreviated notation $\ket{{x}_{|t}}\equiv \ket{x}|_{x=t}$, we have
$\mathbb{S}^{t}=Span\{\ket{\Omega,{\phi_+}_{|t}}\}$. The relevant transition amplitudes can be computed as
\begin{equation}
\begin{split}
\label{test1}
\mathbb{P}_{t',t} (\Omega', \Omega)=\bra{\Omega',{\phi}_{|t'}}\,\Theta (\hat{P}_\phi)\,\hat{\mathbb{P}} \,\Theta (\hat{P}_\phi)\,\ket{\Omega,{\phi}_{|t}}=\frac{\delta (\Omega'-\Omega) \,e^{i|\Omega|\, (t-t')/\beta\hbar}}{{2\alpha |\Omega|}}.
\end{split}
\end{equation}
We then try to solve Eq. \eqref{inner product3} which now takes the form
\begin{eqnarray} 
\big(\Lambda^\dagger_{t} \,\,\mathbb{P}_{t,t} \,\,\Lambda_{t}\big)(\Omega',\Omega)=\int d\bar{\Omega}\,\,\,\Lambda^*_{t}(\bar\Omega,\Omega)\,\Lambda_{t}(\bar\Omega,\Omega')\,(2\alpha |\bar\Omega|)^{-1}=\delta(\Omega',\Omega)
\end{eqnarray}
and has an obvious solution
\begin{eqnarray} 
\Lambda_{t}(\Omega',\Omega)= \sqrt{{2\alpha|{\Omega}|}} \,\delta(\Omega'-\Omega)\,,\,\, \text{or}\,\, \,\hat{{\Lambda}}\equiv \sqrt{{2\alpha|\hat{\Omega}|}}\,\,.
\end{eqnarray}
Next, we examine the condition \eqref{inner product4} using this solution for ${\Lambda}_t$ and \eqref{test1}, and we find
\begin{eqnarray} 
\label{test2}
\big(\Lambda^\dagger_{t} \,\,\mathbb{P}_{t',t} \,\,\Lambda_{t}\big)(\Omega',\Omega)=\delta(\Omega'-\Omega)\,e^{-i|\Omega|\, (t'-t)/\beta\hbar}.
\end{eqnarray}
The result indeed satisfies \eqref{inner product4}, and therefore we have ${\mathbb D}_{t}= {\mathbb D}_{t'>t}\equiv  {\mathbb D}^+$, and so the proposed $t$ serves as a physical time for a Schr\"odinger theory of gravity.

We have thus identified a Schr\"odinger theory with the physical time $t$ associated with the background ${\phi}_+(t)= t$; the theory represents each of the physical states $\bold{\Psi}_{{\mathbb D}^+}\in {\mathbb D}^+$ with the wave functions of the forms $\bold{\Psi}_{{\mathbb D}^+}\big[{\Omega}^{\Lambda}\big](t)$ and $\bold{\Psi}_{{\mathbb D}^+}\big[{p}^{\Lambda}\big](t)$, respectively using the eigenbasis of the complete sets of observables $\hat{\Omega}^{\Lambda}(t)$ and $\hat{p}^{\Lambda}(t)$. 

From the elements of the propagator $\mathcal{T}_{t',t}({\Omega}^{'\Lambda},{\Omega}^{\Lambda})=\big({\Omega}^{'\Lambda}(t')\big|{\Omega}^{\Lambda}(t)\big)= \big(\Lambda^\dagger_{t'} \,\,\mathbb{P}_{t',t} \,\,\Lambda_t\big)(\Omega',\Omega)$ and $\mathcal{T}_{t',t}({p}^{'\Lambda},{p}^{\Lambda})=\big({p}^{'\Lambda}(t')\big|{p}^{\Lambda}(t)\big)= \big(\Lambda^\dagger_{t'} \,\,\mathbb{P}_{t',t} \,\,\Lambda_t\big)(p',p)$, we finally extract the physical Hamiltonian
\begin{eqnarray}
\label{Hamiltonian1}
\hat{\bold H}_{(t)}=\beta^{-1}\big|\hat{\Omega}^{\Lambda}(0)\big|
\end{eqnarray}
governing the evolution of these wavefunctions. Therefore, this Schr\"odinger theory truly has the semiclassical limits of the Hamiltonian dynamics in the reduced phase space $(V,\Omega)$, governed by the effective Hamiltonian ${\bold H}_{(t)}=\beta^{-1}\big|{\Omega} \big|=\beta^{-1}\big|\Delta^{-1/2}p^{3/2}\sin({\frac{\sqrt{\Delta}c}{\sqrt{p}}})\big|$, under the background described by the assigned $\phi(t)=t$ and the dependent $P_\phi(t)={\bold H}_{(t)}$. Actually, this result exactly agrees with the physical Hamiltonian obtained from applying the paradigmatic quantization to this same model \cite{reviewlqc}.

All the established results \cite{improvedlqc}\cite{reviewlqc} in the paradigmatic FRW loop quantum cosmology are thus reproduced through our approach. Particularly,  in our Heisenberg picture, a physical state $\Psi'\in {\mathbb D}^+ $ sharply peaked in the pair $\big(\hat{\Omega}^{\Lambda}\,,\,\hat{p}^{\Lambda})(t_1)$ around a large value of $p$ will remain sharply peaked in $\big(\hat{\Omega}^{\Lambda}\,,\,\hat{p}^{\Lambda})(t)$ for the remaining values of $t$. These peaks define a trajectory of semiclassical cosmology, which agrees with the FRW universe in the large $p$ limits with $\lim_{p\to \infty}{\bold H}_{(t)}=\beta^{-1} |cp|$.  The holonomy corrections as the higher-order terms in the curvature $c$ become important in the small $p$ region, such that the familiar initial singularity is replaced by a cosmic big bounce of the scale factor. Indeed, one can see easily that each trajectory consists of a branch of the collapsing universe with $\partial_t p\propto \cos (b/\hbar)<0$ and another branch of the expanding universe with $\partial_t p \propto \cos (b/\hbar)>0$; the two are joined at the bouncing point where $\sin (b/\hbar)=1$, where the energy density reaches a universal maximal value $\beta^{-1}\Delta^{-1/2}$.

Lastly, let us notice the significance of the proper choice of the background. Observe that the quantum global hyperbolicity, just as in the classical cases, is sensitive to the choice of the background specifying the quantum Cauchy surfaces. Had we chosen the background operator to be $\hat{T}\equiv \hat{\phi}$ instead of $\hat{T}\equiv \hat{\phi}_+$, and with the assigned values of $T(s)=\phi(s)\equiv s $, we would have instead ${\mathbb S}^{s}= Span\big\{\ket{\Omega , {\phi}(s)}\big\}$ corresponding to ${\mathbb D}_s\subset \mathbb{H}$. Going through our algorithm, one would find that Eq. \eqref {inner product3} is again solvable for this background, which thus defines one quantum Cauchy surface $\hat{\Pi}_s:{\mathbb D}_s \to {\mathbb S}^{s}$ for each value in $s$. However, Eq. \eqref {inner product4} in this case is not satisfied, and therefore $ {\mathbb D}_{s'>s}\neq {\mathbb D}_{s}$ and the global hyperbolicity is violated. Therefore, the operator factor $\Theta (\hat{P}_\phi)$, enforcing the classical hyperbolicity in the classical limits, is also necessary for the quantum hyperbolicity at the purely quantum-level.

\subsubsection{ Quantum Modified Klein-Gordon Theory with  $\hat{T}'\equiv \hat{V}_+$ }

We now switch to another type of dynamics not covered by the paradigmatic FRW LQC: the quantum dynamics of the scalar field under the gravitational background. For this we set $\{(\hat{X}_I,\hat{P}_I) \}\equiv \{(\hat{\phi},\hat{P}_{\phi})\}$ and $\{(\hat{X}_{\mu},\hat{P}_{\mu}) \}\equiv \{(\hat{V},\hat{\Omega})\}$. Since the volume scale of the space is often used to label cosmic time for many cosmological models, we will look for a Schr\"odinger theory that has the semiclassical limits of the Hamiltonian theory in the reduced phase space $({\phi},{P}_{\phi})$, with the background sector satisfying $V(\tau)=\tau$ under the physical time $\tau$.

Again, we first look at the form of the classical scalar constraint \eqref{scalar constr} for guidance on the background operator $\hat{T}'$; it is clear that each given set of values for $({\phi},P_{\phi})(\tau)$ and $V(\tau)=\tau$ corresponds to four points on the constraint surface given by the constraint solutions with the four combinations between $\pm\Omega(t)>0$ and $\pm \cos({b}/{\hbar})(t)>0$. Therefore, we are instructed to select only one of these combinations, and here we choose to set $T'\equiv \Theta(\Omega)\Theta(\cos(\frac{b}{\hbar}))\,V\equiv {V}_+$. For a natural quantization of this $T'$, we construct $\hat{T}'(\hat{\phi},\hat{P}_{\phi})\equiv\hat{V}_{+}$ such that
\begin{equation}
\begin{split}
\label{time1}
\ket{{V}_+}\equiv\hat{\Theta} \,\ket{V}\big|_{V=V_+}\,\,\text{with}\,\, \,\hat{\Theta}\equiv\Theta(\hat{\Omega})\,\Theta (\cos(\hat{b}/\hbar)) \,\,;\nonumber
\\
\hat{V}_+\equiv \hat\Theta\, \,\hat{V}\,\,\hat{\Theta}^{\!-\!1}_L=\,\hat{V}\,\,\hat\Theta\,\,+O(\hbar)\,.
\end{split}
\end{equation}

We now set $V_+(\tau)=\tau$ for the proposed physical time $\tau$. Note that the  $\hat{V}_{+}$ and $\hat{V}$ share the same eigenspectrum, so $\tau$ takes values in the discrete set $\{\tau_{n}= 2\pi \gamma l_p^2 \sqrt{\Delta} (1+2n) \}$. Using $\mathbb{S}^{\tau}=Span\{\ket{V_{+|\tau}, P_\phi}\}$, we then calculate the corresponding relevant transition amplitudes via \eqref{FRW riggin map 2} and \eqref{solution form0}, and we obtain:
\begin{eqnarray} 
\label{test5}
\mathbb{P}_{\tau',\tau} (P'_\phi, P_\phi)
&=&\bra{V_{|\tau'}, P'_\phi} \,\Theta (\cos(\hat{b}/\hbar))\,\Theta(\hat{\Omega})\,\,\hat{\mathbb P}\,\,\Theta(\hat{\Omega})\,\Theta (\cos(\hat{b}/\hbar))\,\ket{V_{|\tau}, P_\phi}\nonumber
\\
&=& \delta(P'_\phi -P_\phi)\,\frac{\bra{\Omega_{({P}_{\phi})}}\Theta (\cos(\hat{b}/\hbar))\ket{{V}_{|\tau'}}^*\,\bra{\Omega_{({P}_{\phi})}}\Theta (\cos(\hat{b}/\hbar))\ket{{V}_{|\tau}}}{{2\alpha|{P}_{\phi}|}}\,\,.
\end{eqnarray} 
We then solve Eq. \eqref{inner product3} and find a solution 
\begin{eqnarray} 
\label{test8}
\Lambda_{\tau}(P'_\phi, P_\phi)= \sqrt{\frac{{2\alpha|{P}_{\phi}|}}{|F({P}_{\phi}, V_{|\tau})|^2}} \,\,\delta(P'_\phi- P_\phi)\,,\,\,\text{or}\,\,\,
\,\, \,\hat{{\Lambda}}\equiv \frac{ \sqrt{2\alpha|\hat{P}_{\phi}|}}{|F(\hat{P}_{\phi}, \hat{V}_+)|}\,\,,\nonumber
\\
\text{with}\,\,\,F({P}_{\phi}, V)\equiv \bra{\Omega_{({P}_{\phi})}}\Theta (\cos(\hat{b}/\hbar))\ket{V}.
\end{eqnarray}
Next, we examine the condition \eqref{inner product4} using \eqref{test5} and \eqref{test8}, and we find
\begin{eqnarray} 
\label{test9}
\big(\Lambda^\dagger_{\tau'} \,\,\mathbb{P}_{\tau',\tau} \,\,\Lambda_{\tau}\big)(P'_\phi, P_\phi)
=\delta(P'_\phi- P_\phi)\,\,e^{-i\left[\bar{\theta}_+\left( {P}'_{\phi}\,,\,\tau'\right) -\bar{\theta}_+\left({P}_{\phi}\,,\,\tau\right)\right]}\,\,,\,\, \text{with} \,\, e^{-i \bar{\theta}_+\left({P}_{\phi}\,,\,\tau\right)}\equiv \frac{F({P}_{\phi}, V_{|\tau})}{|F({P}_{\phi}, V_{|\tau})|}.\nonumber\\
\end{eqnarray}
and the condition \eqref{inner product4} is indeed satisfied. So, we have ${\mathbb D}_{\tau}= {\mathbb D}_{\tau'>\tau}\equiv \tilde{\mathbb D}^+$, and this $\tau$ serves as a physical time for a Schr\"odinger theory of the scalar field.

We have thus identified a Schr\"odinger theory with the physical time $\tau$ associated with the background ${V}_+(\tau)= \tau$; the theory represents each of the physical states $\bold{\Psi}_{{\tilde{\mathbb D}}^+}\in {\tilde{\mathbb D}}^+$ with the wave functions of the form $\bold{\Psi}_{\tilde{\mathbb D}^+}\big[{P}_{\phi}^{\Lambda}\big](\tau)$ and $\bold{\Psi}_{\tilde{\mathbb D}^+}\big[{\phi}^{\Lambda}\big](\tau)$, respectively, using the eigenbasis of the complete sets of observables $\hat{P}^{\Lambda}_{\phi}(\tau)$ and $\hat{\phi}^{\Lambda}(\tau)$. 

We now have the elements of the propagator $\mathcal{T}_{\tau',\tau}({P}_{\phi}^{'\Lambda},{{P}_{\phi}}^{\Lambda})=\big({P}_{\phi}^{'\Lambda}(\tau')\big|{P}_{\phi}^{\Lambda}(\tau)\big)= \big(\Lambda^\dagger_{\tau'} \,\,\mathbb{P}_{\tau',\tau} \,\,\Lambda_\tau\big)({P}_{\phi}',{P}_{\phi})$ and $\mathcal{T}_{\tau',\tau}({\phi}^{'\Lambda},{\phi}^{\Lambda})=\big({\phi}^{'\Lambda}(\tau')\big|{\phi}^{\Lambda}(\tau)\big)= \big(\Lambda^\dagger_{\tau'} \,\,\mathbb{P}_{\tau',\tau} \,\,\Lambda_\tau\big)(\phi',\phi)$.  Defined with a discretized notion of time, the propagator takes the form
 \begin{eqnarray}
\label{propagator}
\hat{\mathcal{T}}_{\tau_{n'},\tau_n}= e^{-\frac{i}{\hbar}  \sum_{m=n}^{n'} \Delta\tau\, \hat{\bold H}_{(\tau_m)}}\,;\,\,   \hat{\bold H}_{(\tau_n)}= \frac{\hbar}{\Delta\tau} \left[ \bar{\theta}_+\left( \hat{P}_{\phi}^{\Lambda}(\tau_{0})\,,\,\tau_{n+1}\right) - \bar{\theta}_+\left(\hat{P}_{\phi}^{\Lambda}(\tau_{0})\,,\,\tau_{n}\right)\right].
\end{eqnarray}
where $\Delta\tau \equiv \tau_{m+1}- \tau_{m}= 4\pi \gamma l_p^2 \sqrt{\Delta}$. The corresponding physical Hamiltonian $\hat{\bold H}_{(\tau)}$ can thus be obtained by the values of $\bar{\theta}_+\left( {P}_{\phi}, \tau\right)$ already obtained analytically, which according to \eqref{semiclass4} satisfies
\begin{eqnarray}
\label{theta limit}
 \bar{\theta}_+\left( {P}_{\phi}, \tau\right)
&=&{\theta}_+( \Omega_{({P}_{\phi})}, \tau)+O(l_p^2)\nonumber
\\
&=&\left[\frac{\gamma}{3\beta^2\hbar}\left(\Omega\ln \bigg|\tan\frac{b_{+(\Omega_{({P}_{\phi})},\tau)}}{2\hbar}\bigg|-\frac{b_{+(\Omega_{({P}_{\phi})},\tau)}\,\tau}{\sqrt{\Delta}}\right) +\frac{\pi}{4}\right] +O( l_p^2).
\end{eqnarray}
Thus the physical Hamiltonian satisfies
\begin{eqnarray}
\label{Hamiltonian3 limit}
\hat{\bold H}_{(\tau)}
= \hbar \frac{\partial}{\partial \tau}\,{\bar\theta}_+(\hat{P}_{\phi}^{\Lambda}(\tau_{0}) , \tau)+O(l_p^2\sqrt{\Delta}) 
=\frac{-\gamma}{3\beta^2\hbar\sqrt{\Delta}}{b}_+(\Omega_{(\hat{P}_{\phi}^{\Lambda}(\tau_{0}))},\tau)+O( l_p^2) .
\end{eqnarray}
Note that this Hamiltonian governs the quantum theory of the scalar field, which has the quantum fluctuations of $O(\hbar)$, and so the much smaller error term of $O( l_p^{2})$ may be thought of as a part of the quantum gravitational effect which we may ignore here. Further, for a given moment $\tau=V(\tau)$ we can introduce a physical state of which the energy is bounded by a dimensionless parameter $\epsilon$, and it is given by
 \begin{eqnarray}
\label{low}
\bold{\Psi}^{(\tau,\epsilon)}_{\tilde{\mathbb D}^+}\equiv \int^{\epsilon\,\frac{\tau}{\sqrt{\Delta}\beta}}_{-\epsilon\,\frac{\tau}{\sqrt{\Delta}\beta}} \,\,d {P}_{\phi} \,\,\bold{\Psi}^{(\tau, \,\epsilon)}_{\tilde{\mathbb D}^+}\big[{P}_{\phi}^{\Lambda}\big](\tau)\,\, \,\big|{P}_{\phi}^{\Lambda}(\tau)\big)\,.\,\,
 \end{eqnarray}
According to \eqref{relation} this state satisfies
 \begin{eqnarray}
\frac{{b}_+(\Omega_{(\hat{P}_{\phi}^{\Lambda}(\tau_{0}))},\tau)}{\hbar} \bold{\Psi}^{(\tau,\epsilon)}_{\tilde{\mathbb D}^+}= \sin\left(\frac{{b}_+(\Omega_{(\hat{P}_{\phi}^{\Lambda}(\tau_{0}))},\tau)}{\hbar}\right)\,\,\bold{\Psi}^{(\tau,\epsilon)}_{\tilde{\mathbb D}^+} +O(\epsilon^2)=\frac{\beta\sqrt{\Delta}|\hat{P}_{\phi}^{\Lambda}(\tau_{0})|}{\tau}\,\,\bold{\Psi}^{(\tau,\epsilon)}_{\tilde{\mathbb D}^+} +O(\epsilon^2).\nonumber \\
\end{eqnarray}
It is then straightforward to check that we have
\begin{eqnarray}
\label{low2}
\hat{\bold H}_{(\tau')}\,\bold{\Psi}^{(\tau',\epsilon)}_{\tilde{\mathbb D}^+}=\sqrt{\frac{1}{12\pi G}}\,\frac{\,\,|\hat{P}_{\phi}^{\Lambda}(\tau_{0})|}{\tau}\,\,\bold{\Psi}^{(\tau,\epsilon)}_{\tilde{\mathbb D}^+} +O(\epsilon^2)+O(l_p^2).
\end{eqnarray}

Now we can see the following. The semiclassical limits of this Schr\"odinger theory is a reduced phase space Hamiltonian theory governed by ${\bold H}_{(\tau)}({P}_{\phi},\phi)$. Each moment of time $\tau$ is associated with a growing energy scale, above which the scalar field starts detecting the quantum nature of the gravitation background.  Indeed, for a low-energy state at $\tau$, given by $\epsilon\ll 1$, the reduced phase space Hamiltonian becomes $\sqrt{\frac{1}{12\pi G}}\,\frac{\,\,|{P}_{\phi}|}{\tau}$, which is for the standard homogeneous Klein-Gordon theory in the smooth FRW universe, under the background with the assigned $V(\tau)= p^{3/2}(\tau)=\tau$ and the dependent $\Omega(\tau)=p(\tau)\,c(\tau)=\beta |{P}_{\phi}(\tau)|$. The high-energy states with $\epsilon >1$ at the moment $\tau$ would receive holonomy corrections of $O(\epsilon^2)$, in two different ways. First, the dispersion relation of the modes is drastically corrected, leading to the ultraviolet cutoff in energy as can be seen from the boundedness of ${b}_+(\Omega_{({P}_{\phi}},\tau)/\hbar$. Second, the gravitation background also deviates from the smooth FRW spacetime, now with the assigned $V(\tau)= p^{3/2}(\tau)=\tau$ and the dependent $\big|\Delta^{-1/2}p^{3/2}(\tau)\sin({\frac{\sqrt{\Delta}c(\tau)}{\sqrt{p(\tau)}}})\big|=\beta |{P}_{\phi}(\tau)|$.

Lastly, let us again observe that the exact form of the quantum background $\hat{T}'$-- which shapes the precise details of the above dynamics-- interlocks tightly with the demand of the quantum unitarity. Without using the guiding principle in the beginning, we may have instead chosen $\hat{T}'\equiv \hat{V}$ and set $T'(\rho)=V(\rho)\equiv\rho$ with $\rho$ as a proposed physical time. One can check that Eq. \eqref {inner product3} is solvable for this background. However, the condition \eqref {inner product4} again fails to hold for the solutions. Moreover, when following the guidance, we may have instead used $\hat{T}\equiv \hat{\Theta'}\,\hat{V}\, \hat{\Theta}^{'-1}_L$, with $\hat{\Theta'}\equiv\Theta (\cos(\hat{b}/\hbar))\Theta(\hat{\Omega})$. Yet this choice will also fail the condition \eqref {inner product4} of unitarity, which then asks us to put $\Theta (\cos(\hat{b}/\hbar))$ to the right of $\Theta(\hat{\Omega})$ for the $\hat{\Theta}$ that gives our particular $\hat{T}'$.

\subsubsection{ Transformation between the Reference Frames}

A physical state $\bar{\Psi}_{{\mathbb D}^+\cap \tilde{\mathbb D}^+}\in {\mathbb D}^+\cap \tilde{\mathbb D}^+$ is described by both of the two Schr\"odinger theories we have just derived.  It is then important to study the relation between the two descriptions.

Let us first compute the matrix $\big({P}_{\phi}^{\Lambda}(\tau)\big|{\Omega}^{\Lambda}(t)\big)$ involving the relevant transition amplitudes between $\mathbb{S}^{\tau}$ and $\mathbb{S}^{t}$; according to \eqref{eigenbasis}, it is given by
\begin{eqnarray}
\label{trans amp4}
 \big({P}_{\phi}^{\Lambda}(\tau)\big|{\Omega}^{\Lambda}(t)\big)
&=&\int d\bar{\Omega}d\bar{P}_{\phi} \,\Lambda_{\,\bar{\Omega},\Omega,t}\,\Lambda^*_{\,\bar{P}_{\phi},P_{\phi},\tau}\,\bra{{\bold{V}}_+(\tau), \bar{P}_{\phi}}\hat{\mathbb{P}}\ket{\bar{\Omega},\phi_+(t)}\nonumber\\
&=& \Theta({P}_{\phi}^{\Lambda}(\tau)) \Theta({\Omega}^{\Lambda}(t)) \,e^{-i [\,{P}_{\phi}^{\Lambda}(\tau)\, t\,-\,\bar{\theta}_+({P}_{\phi}^{\Lambda}(\tau),\tau) \,]/\hbar}\delta({P}_{\phi}^{\Lambda}(\tau)-\beta^{-1}{\Omega}^{\Lambda}(t)).
\end{eqnarray}
The subspace ${\mathbb D}^+\cap \tilde{\mathbb D}^+$, where ${\Omega}^{\Lambda}(t)>0$ and ${P}_{\phi}^{\Lambda}(\tau)>0$ are satisfied, is clearly where this matrix becomes unitary. Related by this unitary transformation, the two wave functions $\bar{\Psi}_{{\mathbb D}^+\cap \tilde{\mathbb D}^+}[{P}_{\phi}^{\Lambda}](\tau)$ and $\bar{\Psi}_{{\mathbb D}^+\cap \tilde{\mathbb D}^+}[{\Omega}^{\Lambda}](t)$ describe $\bar{\Psi}_{{\mathbb D}^+\cap \tilde{\mathbb D}^+}$ in the two quantum reference frames associated with the backgrounds ${\phi}_+(t)= t$ and ${V}_+(\tau)=\tau$.

To study the semiclassical limits of this transformation, we pick a state that is semiclassical at the moment of time $t_0$, and then study its description in the other frame at the moment of time $\tau_0$. We first choose the physical state $\bar{\Psi}_{{\mathbb D}^+\cap \tilde{\mathbb D}^+}$ to be sharply peaked in the canonical conjugate observables $\left(\frac{\hat{b}}{\sqrt{\Delta}\hbar}\,, \,\,(2\pi \gamma l_p^2 \sqrt{\Delta})\hat{v}\right)^{\Lambda}(t)=\left( \hat{c}\,\hat{p}^{-1/2}\,,\,\, \hat{p}^{2/3}\right)^{\Lambda}(t)\equiv (\hat{P}_V\,,\,\,\hat{V})^{\Lambda}(t)$ around some values $({P}_{V_0}, V_0)$ corresponding to $(b_0/\hbar \in I_+\,,\,\,v_0)$. Since we have $[\hat{P}_V^{\Lambda}(t)\,,\,\,\hat{V}^{\Lambda}(t)]=4\pi\gamma  G \hbar$,  the wave packet has the width of $\sigma\sim  l_p$. We denote such a wave packet state as $ |({P}_{V_0}, V_0, \sigma)^{\Lambda}(t_0)\big)$ and express it as ($z$ being an normalization constant)
\begin{eqnarray}
\label{coherent state}
\bar{\Psi}_{{\mathbb D}^+\cap \tilde{\mathbb D}^+}
&\equiv&\big|({P}_{V_0}, V_0, \sigma)^{\Lambda}(t_0)\big)
\equiv z\int_{I_+} db \,\,e^{-(b-b_0)^2/\Delta \hbar^2\sigma^2}\, \big({b}^{\Lambda}(t_0)\big|{v_0}^{\Lambda}(t_0)\big)\,\,\big|{b}^{\Lambda}(t_0)\big)\,\,\,\, \nonumber
\\
&& \text{with}\,\,\,\,\big({b}^{\Lambda}(t_0)\big|{v_0}^{\Lambda}(t_0)\big)=\frac{1}{\sqrt{2\pi\hbar}}e^{i\,b \, v_0/2\hbar}.
\end{eqnarray}
To use the transformation formula \eqref{trans amp4}, we make the following change of basis:
\begin{eqnarray}
\label{coherent state 2}
\big|({P}_{V_0}, V_0, \sigma)^{\Lambda}(t_0)\big)
&=& \int_0^{\infty} d\Omega  \,\,\big|\Omega^{\Lambda}(t_0)\big)\big(\Omega^{\Lambda}(t_0)\big|\,\cdot\,\big|({P}_{V_0}, V_0, \sigma)^{\Lambda}(t_0)\big)\nonumber
\\
&=&z\int_0^{\infty} d\Omega \,\,\int_{I_+} db \,\,e^{-(b-b_0)^2/\Delta \hbar^2\sigma^2}\, \big(\Omega^{\Lambda}(t_0)\big|{b}^{\Lambda}(t_0)\big)\big({b}^{\Lambda}(t_0)\big|{v_0}^{\Lambda}(t_0)\big)\,\, \,\,\big|\Omega^{\Lambda}(t_0)\big)\nonumber
\\
&=&z\int_0^\infty d\Omega \,\,e^{-[b_{+(\Omega, V_0)}-b_{+(\Omega_0, V_0)}]^2/\Delta \hbar^2\sigma^2}\, B_+(\Omega,V_0)\,e^{ -i \,\theta_+(\Omega,V_0)}\,|\Omega^{\Lambda}(t_0))+O(l_p^2).\nonumber
\\
\end{eqnarray}
In the last identity, we set $b_0\equiv b_{+(\Omega_0, V_0)}$ and apply the stationary phase approximation given in \eqref{semiclass5}, with the $f(b)$ being the Gaussian distribution.

Introducing the quantities $\beta {P_{\phi}}_0\equiv\Omega_0$, $\phi_0\equiv t_0$, and $V_0 \equiv \tau_0$ and recalling the fact that $\theta_+(\Omega_{(P_\phi)},V_0)=\bar{\theta}_+(P_\phi,V_0)+O(l_p^2)$, we now use \eqref{trans amp4} to perform the frame transformation and obtain the description of the state at the moment $\tau_0$ as
\begin{eqnarray}
\label{coherent trans}
\big|({P}_{V_0}, V_0, \sigma)^{\Lambda}(t_0)\big)
&=&z\beta \int_0^\infty dP_{\phi}  \,\,e^{-[\,b_+(\beta P_{\phi}, \tau_0)-b_+(\beta{P_{\phi}}_0, \tau_0)\,]^2/\Delta \hbar^2\sigma^2}\, B_+(\beta P_{\phi},\tau_0)\,\, e^{-i\, P_{\phi}\, \phi_0\,/\hbar } \big|{P}^{\Lambda}_{\phi}(\tau_0)\big)+O(l_p^2)\nonumber
\\
&\equiv& |({P_{\phi}}_0, \phi_0, \sigma')^{\Lambda}(\tau_0)\big)\,.
\end{eqnarray}
As shown above, the physical state is also peaked in the conjugate pair of observables $(\hat{P}_{\phi}, \hat{\phi},)^{\Lambda}(\tau_0)$ around the values $({P_{\phi}}_0, \phi_0)$ with the width denoted as $\sigma'$. The value of $\sigma'$ can be estimated to be 
\begin{eqnarray}
\label{fluct}
\sigma'\sim\sqrt{\Delta}\,\hbar\,\sigma\left[\frac{\partial}{\partial_{P_{\phi}}}b_+(\beta P_{\phi}, \tau_0)\right]^{-1}\big|_{P_{\phi}={P_{\phi}}_0}=\big(\beta^{-1}\tau_0 \cos(b_0/\hbar)\big)\,\sigma.
\end{eqnarray}

We now make the following observations. First, in the limit of $\hbar\to 0$, the above transformation indeed recovers the corresponding classical reference frame transformation from $({P}_{V_0}, V_0, t_0)$ to $({P_{\phi}}_0, \phi_0, \tau_0)$. Second, at the late enough moments with $\tau_0 \cos(b_0/\hbar)> 1$, we have $\sigma'\sim \beta^{-1} \sigma \gg \sigma$. That is, our physical state $\bar{\Psi}_{{\mathbb D}^+\cap \tilde{\mathbb D}^+}$ gives the gravitational evolution as a sharp trajectory, yet it gives a highly quantum evolution for the scalar field. This gives a concrete (though symmetry-reduced) scenario of a quantum field theory in an effective semiclassical spacetime emerging from a fully quantized gravity-matter coupling system. Third, the relation between the magnitudes of the quantum fluctuations in the gravitation and scalar fields is not constant but evolving dynamically. Particularly, as suggested by \eqref{fluct}, in the cases near the big bounce with $\tau_0 \cos(b_0/\hbar)\sim \beta$ the hierarchical relation between the two may even be inverted by the loop corrections. Lastly, the transformation between the quantum reference frames yields not only the widths and expectation values of the wave packets in the two frames but also the valuable details about the shapes of the wave packets. In this example, the Gaussian wave packet state \eqref{coherent state} in $(P_V, V)$ transforms to the non-Gaussian wave packet state in $(P_{\phi},\phi)$ of the form \eqref{coherent trans} with the distortions given by the functions $b_+(\beta P_{\phi}, \tau)$ and $ B_+(\beta P_{\phi},\tau)$.

\subsubsection{ Quantum Cauchy Surfaces, Global Hyperbolicity and Relational Dirac Observables }

We now explicitly write down the ingredients of the Dirac theory that have been implicitly determined by our calculation above. Recall that the rigging map elements define our physical Hilbert space $\mathbb{H}\subset \mathbb{K}^*$ and supply it with the inner products according to \eqref{inner product}. It is then straightforward to check that $\mathbb{H}$ has an orthonormal basis given by $\big\{ \,\,{|\alpha P_\phi|}^{\frac{-1}{2}}\!\bra{\Omega,\,P_\phi}\,;\,\,|\beta P_\phi|\!=\!|\Omega|\,\, \big\}$, which we may denote as either $\big\{|\pm, P_{\phi}\,)\big\}$ or $\big\{|\Omega,\pm\,)\big\}$ since the absolute values of the two arguments are constrained by one another. We have obtained the quantum Cauchy surfaces $\hat{\Pi}_t: {\mathbb D}^+\to {\mathbb S}^{t}$ and $\hat{\Pi}_{\tau}: \tilde{\mathbb D}^+ \to {\mathbb S}^{\tau}$ providing a global hyperbolic foliation to, respectively, the quantum spacetimes in ${\mathbb D}^+$ and $\tilde{\mathbb D}^+$. In reverse, one can apply $\hat{\mathbb{P}}$ to ${\mathbb S}^{t}$ and ${\mathbb S}^{\tau}$ and verify that we have ${\mathbb D}^+= Span\big\{|\Omega, +\,) \big\}$ and $\tilde{\mathbb D}^+= Span\big\{|+, P_{\phi}\,) \big\}$.

According to \eqref{Dirac observ 0}, the relational observables $\big(\hat{\Omega}^{\Lambda}\,,\,\hat{p}^{\Lambda})(t)$ and $ \big(\hat{\phi}^{\Lambda}\,,\,\hat{P}_{\phi}^{\Lambda}\big)(\tau)$ are of the forms
\begin{eqnarray}
\label{observables B}
\big(\hat{\Omega}^{\Lambda}\,,\,\hat{p}^{\Lambda})(t)\bigg|_{\mathbb{D}^+}
&\equiv&\,\, \hat{\mathbb P}\,\,\sqrt{{2|\alpha \hat{\Omega}|}}\,\, \big(\hat{\Omega}\,,\,\hat{p}\big) \,\,\frac{1}{\sqrt{{2|\alpha \hat{\Omega}|}}}\,\, \hat{\Pi}_t\nonumber
\\
\big(\hat{\phi}^{\Lambda}\,,\,\hat{P}_{\phi}^{\Lambda}\big)(\tau)\bigg|_{\tilde{\mathbb{D}}^+}
&\equiv&\,\, \hat{\mathbb P}\,\,\frac{ \sqrt{2\alpha|\hat{P}_{\phi}|}}{|F(\hat{P}_{\phi}, \hat{V}_+)|}\,\, \big(\hat{\phi}\,,\,\hat{P}_{\phi}\big) \,\,\frac{|F(\hat{P}_{\phi}, \hat{V}_+)|}{ \sqrt{2\alpha|\hat{P}_{\phi}|}}\,\, \hat{\Pi}_{\tau}\,.
\end{eqnarray} 
One can further express these in terms of the kinematic complete sets and obtain
\begin{eqnarray}
\label{relational observ explicit}
 ( \hat{\Omega}, \hat{p})^{\Lambda}(t)
&=& \int_{-\infty}^{\infty}\,d\lambda\, \,\,e^{i \lambda\hat{C}/\hbar} \,\,\,\,\sqrt{{2\alpha|\hat{\Omega}|}}\,\,( \hat{\Omega}, \hat{p}) \,\,\frac{1}{\sqrt{{2\alpha|\hat{\Omega}|}}}\,\,\,\Theta(\hat{P}_{\phi}) \,\delta(\hat{\phi}-t)\Theta(\hat{P}_{\phi}) \,\,|\hat{\dot{\phi}}|\,\, \,\,\,e^{-i \lambda\hat{C}/\hbar}\nonumber
\\
( \hat{P}_{\phi}, \hat{\phi})^{\Lambda}(\tau)
&=& \int_{-\infty}^{\infty}\,d\lambda\, \,\,e^{i \lambda\hat{C}/\hbar} \,\,\,\,\frac{ \sqrt{2\alpha|\hat{P}_{\phi}|}}{|F(\hat{P}_{\phi}, \hat{V}_+)|}\,\,( \hat{P}_\phi, \hat{\phi}) \,\,\frac{|F(\hat{P}_{\phi}, \hat{V}_+)|}{ \sqrt{2\alpha|\hat{P}_{\phi}|}} \times\nonumber 
\\
&&\,\,\,\,\,\,\,\,\Theta(\hat{\Omega}) \Theta(\cos \hat{b}/\hbar)\,\,\widehat{\delta({V}-\tau)}\,\,\Theta(\cos \hat{b}/\hbar)\Theta(\hat{\Omega})  \,\,|\hat{\dot{V}}| \,\,\,\,\,e^{-i \lambda\hat{C}/\hbar}\,,
\nonumber\\
\end{eqnarray}
where we have used the notations 
\begin{eqnarray}
\label{def}
\hat{\dot{\phi}}\equiv \hat{P}_{\phi}\,\,;\,\,
\hat{\dot{V}}\equiv \,\frac{\hat{\Omega}}{F^{2}(\hat{\Omega}/\beta, \tau)}\,;\,\,\,
\widehat{\delta({V}-\tau)}\equiv \frac{1}{{\Delta{V}}}\ket{V(\tau)}\bra{V(\tau)}\,
\end{eqnarray}
based on their verifiable semiclassical limits, with the dots denoting the differentiation with respect to the $\lambda$. We can see now these observables are truly quantum relational observables representing the gauge-invariant phase space functions
\begin{eqnarray}
\label{relational observ3}
({\Omega}, {p})(t)
&=& \int_{-\infty}^{\infty}\,\,d\lambda \,\delta({\phi}(\lambda)-t) \,\,|{\dot{\phi}}(\lambda)|\,\,\Theta({P}_{\phi}(\lambda))\,\,({\Omega}(\lambda), {p}(\lambda))  \, \nonumber
\\
({P}_{\phi}, {\phi})(\tau)
&=& \int_{-\infty}^{\infty}\,\,d\lambda \,\delta({V}(\lambda)-\tau) \,\,|{\dot{V}}(\lambda)|\,\,\Theta(\Omega(\lambda))\,\Theta(\,\cos (\,c\sqrt{\Delta/p}\,)\,)\,({P}_{\phi}(\lambda), {\phi}(\lambda)) 
\end{eqnarray}
where $\lambda$ serves as the parameter of the group generated by $\mathcal{C}$. In this form, the quantum Cauchy surfaces serve as fundamental objects in providing the relational observables faithfully representing $\big(\hat{\Omega}\,,\,\hat{p}\big)$ and $\big(\hat{\phi}\,,\,\hat{P}_{\phi}\big)$. This then allows the emergence of the Schr\"odinger theories from the timeless Dirac theory.

Finally, the full physical Hilbert space $\mathbb{H}={\mathbb D}^+\cup{\mathbb D}^-=\tilde{\mathbb D}^+\cup\tilde{\mathbb D}^-$ contains both of the ``$+$" and ``$-$" branches of the Schr\"odinger theories. This means that every quantum state in either branch of the quantum gravity theories, or in either branch of the deformed Klein-Gordon theories, represents exactly one physical state in $\mathbb H$. Further, a physical state in any one of the subspaces $\{{\mathbb D}^+ \cap \tilde{\mathbb D}^+, {\mathbb D}^- \cap \tilde{\mathbb D}^-, {\mathbb D}^+ \cap \tilde{\mathbb D}^-, {\mathbb D}^- \cap \tilde{\mathbb D}^+\}$ is describable by both types of dynamics.

\section{Summary and Conclusion}

Based on the foundation of the previous works \cite{CY0}\cite{CY1}\cite{CY2}, we proposed an algorithm of transforming the relevant elements of a well-defined Einstein-Hilbert path integral $\hat{\mathbb{P}}:\mathbb{K} \to \mathbb{K}^*$, into the Schr\"odinger propagator under each valid notion of physical time. The only input of the algorithm is the relevant matrix elements of $\hat{\mathbb{P}}$ corresponding to the transition amplitudes between the eigenspaces $\{\mathbb{S}^t\subset \mathbb{K}\}$ of the quantum background fields $\hat{T}$, the values $T(t)=t$ of which mark the moments of the proposed physical time $t$. 

The operator $\hat{\Lambda}$ solved from these matrix elements provides the transformation from the timeless Fadeev-Popov path integral into the reduced phase space path integrals under all viable notions of physical time. Further, there is a generalized Heisenberg picture unifying the resulting Schr\"odinger theories into one timeless theory, which turns out to be the canonical theory with the physical Hilbert space $\mathbb{H}\subset\mathbb{K}^*$ from the image of $\hat{\mathbb{P}}$ acting as the rigging map. In this generalized Heisenberg picture, each moment of physical time is a quantum Cauchy surface $\hat{\Pi}_t: \mathbb{D}_t \to \mathbb{S}^t$ supplying $\mathbb{S}^t$ as a faithful representation for the corresponding physical states $\mathbb{D}_t \subset\mathbb{H}$. Consequentially, a complete set of observables in $\mathbb{D}_t$ can be induced by a complete set of self-adjoint operators in the ``quantum reduced phase space" $\mathbb{S}^t$, and these are the elementary quantum relational observables defined with a specified background value $T=t$. Altogether, we have formulated an exact notion of quantum reference frames for the timeless canonical quantum gravity, through which the Schr\"odinger theories under various notions of physical time can be calculated from just the elements of $\hat{\mathbb{P}}$.

In this paper, we have demonstrated its application to the FRW loop quantum cosmology with a massless Klein-Gordon scalar field. From the transition amplitudes $\hat{\mathbb{P}}$ of the model, we derived two interesting Schr\"odinger theories in two quantum reference frames: the quantum gravidity theory in the reference frame specified by the scalar field background, and the modified Klein-Gordon theory in the reference frame specified by the gravitation background. The descriptions in the different reference frames reveal a wider and deeper view of the quantum geometric effects introduced in the loop quantization, and the effects manifest in different forms of quantum dynamics. Viewed in the two frames, the quantum geometry causes the big bounce resolution of the initial singularity in the gravitation evolution, the time-dependent high-energy deformation of the Klein-Gordon field evolution, and also the time-dependent correlation between the quantum fluctuation scales in the two dynamics. 

Under the development of increasingly realistic and sophisticated quantum cosmological models \cite{perturbationqc}\cite{inhomogeneouslqc}\cite{inhomogeneouslqc2}\cite{bhqc}\cite{bhbounce1}, we are starting to confront the cases in which the solution space of the quantum constraints is not well understood. That means we are no longer given a clear characterization of the physical states, such as the mentioned conserved currents, as the tools for extracting the Schr\"odinger dynamics. This situation calls for a fundamental and computable approach to the emergence of Schr\"odinger dynamics. Since in most of these models the transition amplitudes ${\mathbb{P}}$ are well defined and can be calculated perturbatively, the quantum Cauchy surfaces are predominantly meaningful among them. Therefore, our algorithm may fulfill the demand to extract these model's dynamics in a universally manner, describing them using the elementary relational Dirac observables. Particularly, the algorithm can be carried out order by order within a perturbation scheme for the transition amplitudes, leading to the corresponding perturbation expansion of the $\hat{\Lambda}$; Meanwhile, by construction, the elementary observable algebra \eqref{Dirac observ 1} will remain order-independently exact, so that the perturbation becomes that of the propagator in an ordinary Schr\"odinger theory. 

Viewed from the full theories' perspective, our approach is especially meaningful to loop quantum gravity \cite{lqg1}\cite{lqg2}\cite{Dirac1}. The theory is built from a robust kinematic Hilbert space $\mathbb{K}$ with a basis given by ``spin-network states," each of which is defined with a colored graph consisting of oriented edges and vertices in the spatial manifold. This space $\mathbb{K}$ is equipped with a complete set of self-adjoint gravitation and matter operators, built from the conjugate pairs of flux and holonomy variables associated with the graphs of the states. The flux-holonomy quantum algebra offers a compelling background independent description of the system: the gravitational coloring describes the spatial quantum geometry \cite{area}\cite{volume}, with the quanta of the area and volume, respectively, carried by the edges and nodes, and the matter coloring describes the matter content's flux or holonomy excitations \cite{matter} upon the spatial quantum geometry. It is thus desirable for this flux-holonomy algebra to survive in the physical level, so it can take fundamental roles in the dynamics of the Universe. Also, the concrete formulations \cite{projector1}\cite{projector2}\cite{projector3} of $\hat{\mathbb{P}}$ in loop quantum gravity advance remarkably in both covariant and canonical formulations. In the spin-foam models \cite{projector1}\cite{projector3}, each amplitude is formulated as an expansion of the sum over the ``spin foams" connecting the initial and final spin-network states; each spin foam denotes a product of vertex and face amplitudes under analogous Feymann rules. In the canonical approach, the successful construction of the quantum constraints $\{\hat{C}_\mu \}$ acting on the spin-network states has also led to a perturbative construction \cite{projector1}\cite{projector2} of the rigging map operator $\hat{\mathbb{P}}$, making it possible to calculate the elements and derive a spin-foam model from the canonical theory. These developments call for a satisfactory deployment of the $\hat{\mathbb{P}}$ to obtain the physical dynamics. As demonstrated in our earlier works \cite{CY0}\cite{CY1}\cite{CY2}, our proposal may offer such a universal way to cast the full theory into Schr\"odinger theories in variable reference frames, the instantaneous physical degrees of freedom of which are explicitly captured by the spin-network states lying in the quantum Cauchy surfaces, and so the flux-holonomy algebra of quantum geometry may become the fundamental algebra of the observables. 

We thus invite the reader to make use of this algorithm to its full conceptual and practical potential.

\section{Acknowledgements}

We benefited greatly from our numerous discussions with Wojciech Kami\'nski. This work was supported by the grant of the Polish Narodowe Centrum Nauki Grant No. 2011/02/A/ST2/00300.

\end{document}